\documentclass{article}

\usepackage{amsmath}
\usepackage{PRIMEarxiv}
\usepackage{mathrsfs}
\usepackage{listings}
\usepackage{longtable}
\usepackage{tabularx}
\usepackage{forloop}
\usepackage{etoolbox}
\usepackage{multirow}
\usepackage{makecell}
\usepackage{subcaption}
\usepackage{float}
\usepackage{calc}
\usepackage{wrapfig,lipsum,booktabs} 
\usepackage{xtab} 
\usepackage[dvipsnames]{xcolor}

\usepackage[utf8]{inputenc} 
\usepackage[T1]{fontenc}    
\usepackage{hyperref}       
\usepackage{url}            
\usepackage{booktabs}       
\usepackage{amsfonts}       
\usepackage{nicefrac}       
\usepackage{microtype}      
\usepackage{lipsum}
\usepackage{fancyhdr}       
\usepackage{graphicx}       
\graphicspath{{media/}}     

\title{Sentiment-Aware Mean-Variance Portfolio Optimization for Cryptocurrencies}

\author{
  Qizhao Chen \\
  Graduate School of Information Science \\
  University of Hyogo \\
  Kobe, Japan\\
  \texttt{af24o008@guh.u-hyogo.ac.jp} \\
}

\begin{document}
\maketitle

\begin{abstract}
Cryptocurrency markets are highly volatile and influenced by both price trends and market sentiment, making effective portfolio management challenging. This paper proposes a dynamic cryptocurrency portfolio strategy that integrates technical indicators and sentiment analysis to enhance investment decision-making. Market momentum is captured using the 14-day Relative Strength Index (RSI) and Simple Moving Average (SMA), while sentiment signals are extracted from news articles with VADER and further validated using the Google Gemini large language model. These signals are incorporated into expected return estimates and used in a constrained mean-variance optimization framework. Backtesting across multiple cryptocurrencies shows that the integrated approach outperforms traditional benchmarks, including momentum strategy, Bitcoin Long-Short strategy, and an equal-weighted portfolio, achieving stronger risk-adjusted returns and more consistent cumulative growth. Furthermore, comparing the sentiment-only and technical-only strategies shows that incorporating sentiment information alongside technical indicators can lead to more consistent performance gains. However, the strategies exhibit substantial drawdowns that coincide with known periods of market stress, indicating that additional risk-management components are required to improve stability.
\end{abstract}


\section{Introduction}\label{sec1}

Cryptocurrencies have become a popular asset class in recent years, attracting both retail and institutional investors. Due to their high volatility and speculative nature, effective portfolio optimization methods are crucial for managing risk and maximizing returns. Traditional financial models, such as mean-variance optimization (MVO), have been widely used for portfolio allocation in conventional markets. However, applying these methods directly to cryptocurrency markets is challenging due to their unique characteristics, including rapid price swings, lack of regulation, and the strong influence of public sentiment.

Investor behavior in the financial market such as cryptocurrency market is often driven by emotions and reactions to news events, social media trends, and global uncertainty~\cite{10.1145/3649451}. Unlike traditional assets, the value of cryptocurrencies can be significantly affected by market sentiment~\cite{data10040050}. This opens up the possibility of enhancing financial models by incorporating sentiment indicators. In particular, large-scale financial news platforms such as CoinDesk provide real-time updates that reflect the mood of the market. When used appropriately, these signals can offer additional insights beyond historical prices and volatility.

Recent advancements in large language models (LLMs) such as GPT and Gemini have opened new opportunities for leveraging natural language understanding in financial decision-making~\cite{10973330}. These models can process unstructured textual data, including news headlines, social media posts, and analyst reports, with greater contextual awareness than traditional sentiment analysis tools, enabling more nuanced interpretations of market mood. By evaluating not only the polarity of sentiment but also the credibility of sources, the presence of speculative language, and the broader market context, LLMs can generate investment recommendations that are both sentiment-aware and reasoned~\cite{chen2025sentimentawarestockpriceprediction}. Integrating LLM-driven insights into portfolio optimization frameworks allows for dynamic adjustments to expected returns and risk assessments, potentially improving responsiveness to rapidly changing market conditions that characterize cryptocurrency trading. 

This paper proposes a sentiment-enhanced mean-variance portfolio optimization framework for cryptocurrency investment. The method combines traditional MVO with technical and sentiment-based signals. Technical indicators such as Relative Strength Index (RSI) and Simple Moving Average (SMA) are used to detect overbought or oversold conditions, while news-based sentiment scores help capture market mood. A simple signal-weighting mechanism adjusts the expected return of each asset based on these indicators. The portfolio is then re-optimized daily using a sliding window approach to adapt to changing market dynamics.

We conduct empirical backtesting on a set of major cryptocurrencies over a five-year period. The proposed strategy achieves a higher cumulative return and higher Sharpe ratio compared to baselines such as Bitcoin investment and equal-weighted cryptocurrency portfolio. These findings suggest that incorporating sentiment and technical signals into classical portfolio models can provide meaningful improvements, especially in highly speculative and reactive markets like cryptocurrencies. 

The main contributions of this work are as follows:

\begin{enumerate}
    \item \textbf{Sentiment-Enhanced Portfolio Optimization Framework} -- This study proposes a dynamic cryptocurrency portfolio optimization strategy that integrates technical indicators with sentiment signals, enabling more adaptive asset allocation in volatile markets.
    
    \item \textbf{Integration of LLM-Based Sentiment Verification} -- This study extends traditional sentiment scoring by incorporating LLM analysis to verify VADER-generated sentiment scores and provide context-aware investment recommendations.
\end{enumerate}

The rest of the paper is organized as follows. Section 2 reviews the relevant literature. Section 3 outlines the proposed methodology, followed by the presentation of experimental findings in Section 4. Section 5 offers a detailed discussion on validating the VADER sentiment scores using an LLM. Section 6 concludes the study.

\section{Related Work}\label{sec2}

Portfolio optimization has been a central topic in finance for decades. The classical mean-variance model introduced by Markowitz remains a foundational approach for asset allocation. Many studies have applied this framework to various markets, including stocks, bonds, and commodities. In recent years, researchers have explored how this model performs in cryptocurrency markets, where price movements are often more erratic. For example, Serban and Vrinceanu ~\cite{math13101693} propose the Mean–Variance–Entropy (MVE) model, which incorporates Tsallis entropy into traditional mean-variance optimization to enhance portfolio diversification and reduce concentration risk. Applied to a cryptocurrency portfolio, the MVE model demonstrates improved diversification and robustness in volatile markets, albeit with a slightly lower Sharpe ratio compared to the classic approach.  Brauneis and Mestel~\cite{BRAUNEIS2019259} apply the Markowitz mean-variance framework to evaluate the risk-return characteristics of cryptocurrency portfolios using daily data from 2015 to 2017. The findings indicate that diversification across cryptocurrencies reduces risk compared to single-asset investments and benchmarks, with the equally weighted (1/N) portfolio outperforming most mean-variance optimized portfolios and individual cryptocurrencies in terms of Sharpe ratio and certainty equivalent returns, even after accounting for transaction costs.  Gaied Chortane and Naoui ~\cite{jrfm18020077} evaluate whether the traditional mean-variance framework remains effective by replacing it with portfolio optimization methods based on relative Tsallis entropy and mutual information to improve risk management and diversification in complex financial markets. Using daily returns from the S\&P 500 and Bitwise 10 cryptocurrency indices (2019–2024) and analyzing extreme shocks through 2020, the study finds that entropy-based models outperform conventional approaches, particularly in managing extreme risks and capturing nonlinear asset dependencies.  Alidaee et al.~\cite{10926863} evaluate various portfolio models based on Modern Portfolio Theory, including discrete Markowitz, dynamic portfolio approaches, and naive diversification, in US stock and cryptocurrency markets. The results show that discrete Markowitz and optimal dynamic portfolios outperform other models and ETFs with faster computation times, while data frequency (weekly to minute-level) has minimal impact on portfolio selection, highlighting advances in computational efficiency for large, high-frequency datasets.  Slusarczyk and Slepaczuk~\cite{Slusarczyk2025} examine whether incorporating forecasted stock returns from ARIMA-GARCH and XGBoost models into a mean-variance portfolio can improve performance compared to using historical returns alone. By testing DJIA stocks from 2007 to 2022, the results show that forecast-based portfolios can enhance selection within the Markowitz framework, though careful tuning of model parameters is essential, and challenges arise from the large number of model fits required.  Sakowski and Wysocki~\cite{Sakowski2024} develop a deep learning-based framework for estimating the variance–covariance matrix in portfolio optimization, using LSTM-RNN, DeepVAR, and GPVAR for one-day-ahead multivariate forecasts of stocks and cryptocurrencies. Results show that LSTM-RNN models generally achieve the highest information ratios and annualized returns, with longer observation windows and less frequent rebalancing improving performance across all methods.

At the same time, there has been growing interest in using alternative data sources for financial modeling~\cite{chen2025novel}. Sentiment analysis has emerged as a useful tool for capturing market psychology. Several studies have shown that investor sentiment, measured through social media posts, news articles, or search trends, can predict short-term price movements in both equity and cryptocurrency markets. For example, Arslan ~\cite{arslan2025bitcoin} investigates the relationship between Twitter sentiment and Bitcoin price movements, proposing a two-network LSTM-based model. One network processes historical price data decomposed via Empirical Mode Decomposition, while the other integrates classified positive/negative Twitter sentiments with market data. The combined predictions outperform state-of-the-art methods in accuracy, demonstrating the value of sentiment analysis for the forecasting of cryptocurrency. Shang~\cite{10947713} develops a CART decision tree model combining 124 technical indicators with Twitter-roBERTa sentiment analysis to forecast Bitcoin price ranges for the next day, using six years of market data and 58 million tweets. Incorporating sentiment improves the average prediction accuracy from 0.56 to 0.62 and significantly increases win rates.  Chen and Kawashima~\cite{10825946} evaluate large language model (LLM)–based news sentiment estimation for stock price forecasting and find that incorporating news sentiment improves forecasting accuracy.  Alghamdi et al.~\cite{Alghamdi2022} investigate the link between investor sentiment on Twitter and cryptocurrency price volatility, using LSTM for price forecasting and SVM/Naive Bayes for sentiment classification. Results show that SVM outperforms NB with over 93\% accuracy for Bitcoin and Ethereum sentiment analysis, and the forecasting model achieves low error rates (MAE: 0.2545, MSE: 0.2528, RMSE: 0.5028). Frohmann et al.~\cite{bdcc7030137} propose a hybrid Bitcoin price prediction approach that combines historical price data with sentiment analysis from microblogs, introducing a fine-tuned BERT model and a novel follower-weighted sentiment scheme. Experiments show that the BERT-based approach with linear regression achieves the best accuracy (MAE: 2.67, RMSE: 3.28), outperforming previous methods, while complex models like LSTM and TCN face overfitting and generalization challenges.

Combining sentiment with portfolio optimization has also been explored in traditional finance. For example, Colasanto et al.~\cite{colasanto2022bert} integrate sentiment scores from Financial Times articles into the Black–Litterman model to enhance stock forecasting and dynamic portfolio optimization. Using a Monte Carlo approach to simulate price paths after news publication, the method demonstrates that incorporating sentiment-based views improves forecasting accuracy compared to ignoring sentiment. Taheripour et al.~\cite{10977954} enhance traditional portfolio optimization by integrating companies’ financial and operational performance, extracted from quarterly reports using the FinBERT model, into adjusted returns, and addressing return uncertainty via fuzzy trapezoidal numbers. Using a credibilistic conditional value at risk (CCVaR) framework with real-world DJIA data, the proposed method balances risk and return while incorporating transaction costs and practical constraints, yielding diversified and robust portfolios. However, in the context of crypto assets, sentiment-aware portfolio construction remains a relatively underexplored area. Existing studies often focus on single-asset forecasting rather than multi-asset allocation.

Technical indicators are also widely used in trading strategies. RSI, SMA, and MACD are among the most common tools for identifying trend reversals or momentum signals. These indicators have been incorporated into rule-based trading systems and machine learning models. For example, Vargas et al.~\cite{8489208} compare two deep learning approaches for predicting daily stock price direction using financial news titles and technical indicators: a hybrid CNN–LSTM model (SI-RCNN) and an LSTM-only model (I-RNN). Results show that financial news plays a key role in improving prediction stability, while different sets of technical indicators have minimal impact on performance. Vincent et al.~\cite{10284633} improve stock market prediction by combining multiple technical indicators with an LSTM neural network, using historical stock data to generate buy/sell signals. While technical analysis alone can be effective in short-term trading, it often lacks the ability to adapt to sudden shifts caused by news or events.

This paper builds on previous work by combining sentiment and technical indicators within a mean-variance optimization framework. The goal is to assess whether these enhancements can improve portfolio performance in the volatile and sentiment-driven world of cryptocurrencies.

\section{Data and Methodology}\label{sec3}

\subsection{Data}
This study utilizes both structured and unstructured data to construct a sentiment-enhanced portfolio optimization model for cryptocurrencies. The structured data primarily includes historical price and volume information of selected cryptocurrencies, which include Bitcoin (BTC), Ethereum (ETH), Cardano (ADA), Binance Coin (BNB), and Ripple (XRP). These five cryptocurrencies are selected due to their high market capitalization, liquidity, and widespread investor attention, ensuring both reliable historical data and abundant sentiment information. They also represent diverse technological foundations and use cases, offering a broader perspective for sentiment-enhanced portfolio optimization.

\textbf{Bitcoin}, introduced in 2009, is the first decentralized cryptocurrency and is widely regarded as a digital store of value. \textbf{Ethereum}, launched in 2015, is a blockchain platform enabling smart contracts and decentralized applications, with Ether (ETH) serving as its native currency. \textbf{Cardano} is a proof-of-stake blockchain emphasizing scalability and sustainability, with \textbf{ADA} used for transactions, governance, and staking. \textbf{Binance Coin (BNB)}, initially a utility token for discounted fees on the Binance exchange, now supports a range of transactions within the Binance ecosystem. \textbf{Ripple (XRP)} is a digital asset optimized for low-cost, rapid cross-border payments between financial institutions.

The data spans from February 14, 2020, to August 6, 2025, with a daily frequency. All price and volume data are obtained using the Crypto Compare API\footnote{https://developers.coindesk.com/}, which provides reliable and comprehensive market data across a wide range of digital assets. All timestamps are converted to Coordinated Universal Time (UTC) to ensure consistency across different exchanges. Table~\ref{tab:crypto_stats} shows the descriptive statistics of the closing prices of all the cryptocurrencies. The set of cryptocurrencies considered in this study is fixed over the sample period, ensuring consistency in portfolio construction.

\begin{table}[htbp]
\centering
\caption{Descriptive statistics of cryptocurrency closing prices}
\label{tab:crypto_stats}
\begin{tabular}{lrrrrr}
\hline
        & BTC & ETH & ADA & BNB & XRP \\
\hline
count   & 2001.00 & 2001.00 & 2001.00 & 2001.00 & 2001.00 \\
mean    & 44053.63 & 2096.51 & 0.65 & 352.68 & 0.80 \\
std     & 28135.05 & 1097.27 & 0.55 & 212.10 & 0.70 \\
min     & 4916.78 & 110.23 & 0.02 & 9.42 & 0.14 \\
25\%    & 22201.44 & 1472.15 & 0.31 & 234.52 & 0.39 \\
50\%    & 38184.86 & 1967.67 & 0.46 & 316.69 & 0.54 \\
75\%    & 60863.56 & 2945.17 & 0.86 & 558.23 & 0.82 \\
max     & 122184.23 & 4810.97 & 2.97 & 844.17 & 3.55 \\
\hline
\end{tabular}
\end{table}

In addition to price-based data, unstructured textual information is incorporated in the form of sentiment extracted from cryptocurrency-related news articles. News headlines and article bodies are collected using the Crypto Compare News API\footnote{https://developers.coindesk.com/documentation/data-api/news\_v1\_article\_list} over the same time period and with matching daily granularity. Each article is mapped to the corresponding cryptocurrency based on its category and content, and daily sentiment scores are computed as the average VADER compound score of all articles for that asset. These articles cover market updates, investor sentiment, regulatory news, and technological developments within the cryptocurrency space. VADER (Valence Aware Dictionary and sEntiment Reasoner)~\cite{Hutto_Gilbert_2014} is a lexicon-based approach. It assigns each article a compound score, a normalized metric ranging from $-1$ (most negative) to $+1$ (most positive), reflecting the overall sentiment tone of the article. The sentiment data is then aggregated by asset and date to match the price data frequency. 

Furthermore, technical indicators are extracted from the price data using the pandas\_ta library. Indicators such as the 14-day Relative Strength Index (RSI) and 14-day Simple Moving Average (SMA) are calculated. The SMA smooths out price fluctuations by averaging the closing prices over the past 14 days, helping to identify price trends. The RSI is a momentum oscillator that measures the speed and magnitude of recent price changes, providing a value between 0 and 100 to indicate whether an asset is overbought or oversold. 

All features, technical indicators, sentiment scores, and price-based variables, are aligned and merged to form a unified feature matrix that serves as input for the portfolio optimization model. The combination of structured and unstructured data allows for a more comprehensive understanding of market dynamics. The risk-free rate used for portfolio optimization is obtained from the U.S. Treasury 3-month bill yield.

To avoid timing mistakes, all cryptocurrency data—both prices and news—is converted to Coordinated Universal Time (UTC), which is a standard time reference used worldwide. This ensures that information from different exchanges and sources is properly aligned. The portfolio is built using only past data up to the end of the training period, and returns are calculated on the following day. This approach helps prevent using information from the future or mixing up data from different time zones.

\subsection{Feature Engineering and Return Adjustment}

To enhance the quality of the expected return estimates used in portfolio optimization, a feature engineering process is applied that integrates both technical indicators and sentiment information. For each cryptocurrency in the portfolio, the closing price series is extracted. Using this series, two widely used technical indicators are computed: the 14-day Relative Strength Index (RSI) and the 14-day Simple Moving Average (SMA). The RSI is a momentum oscillator that produces a value between 0 and 100. A value above 50 typically indicates bullish momentum, while a value below 50 signals bearish momentum.

The SMA is used to assess the trend direction of the asset’s price. Specifically, the latest price is compared against the 14-day SMA to generate a binary trend signal indicating whether the market is currently in an upward or downward phase. In addition to these technical indicators, sentiment information is incorporated through a daily sentiment score derived from textual data. Although sentiment is often relatively stable at short horizons, it is treated here as a time-varying signal to enable the model to respond to fluctuations in news flow and market attention.

These features are integrated through a Ridge regression model, which is estimated within each rolling window to capture the relationship between technical indicators, sentiment dynamics, and next-day returns. A linear specification is adopted to provide a transparent and interpretable aggregation of heterogeneous signals, a common practice in return predictability and factor-based modeling~\cite{10.1093/rfs/hhp063}. Moreover, linear models are well suited for rolling-window estimation, where stability and robustness are essential given the limited sample size and the noisy nature of financial returns~\cite{10.1145/3770177.3770282}. For each day \(t\), a feature vector is constructed as
\[
X_t =
\begin{bmatrix}
\text{RSI}_t \\
\mathbf{1}(P_t > \text{SMA}_t) \\
s_t
\end{bmatrix},
\]
where \(\mathbf{1}(P_t > \text{SMA}_t)\) is the binary SMA trend indicator, and \(s_t\) is the contemporaneous sentiment value. The Ridge regression coefficients \(\hat{\beta}\) are estimated by solving
\[
\hat{\beta}
= \arg\min_{\beta}
\left[
\sum_{t} \left(r_{t+1} - X_t^\top \beta\right)^2
+ \alpha \|\beta\|_2^2
\right],
\]
where \(r_{t+1}\) denotes the next-day return and \(\alpha\) controls the strength of regularization. This formulation provides a stable estimate of how each feature contributes to return predictability, mitigating overfitting while preserving signal interpretability.

After estimating the model for each rolling window, the most recent feature vector \(X_{t}\) is evaluated using the fitted coefficients to obtain a data-driven adjustment term:
\[
\widehat{\epsilon}_{t} = X_t^\top \hat{\beta}.
\]
This quantity represents the return component predicted by the combined technical-and-sentiment signal set.

The final adjusted expected return used in the portfolio optimization step is computed as
\begin{equation}
\mu_{\text{adj, close}}
= \mu_{\text{close}}
+ \widehat{\epsilon}_{t},
\label{eq:mu_adjustment_ridge}
\end{equation}
where \(\mu_{\text{close}}\) is the empirical mean return over the training window, and \(\widehat{\epsilon}_{t}\) embeds the forward-looking signal extracted from the feature-based regression model. This adjustment ensures that the optimization incorporates not only historical trend information but also the contemporaneous market conditions reflected in momentum, trend, and sentiment dynamics.

\subsection{Model Setup}

Mean-variance portfolio optimization, originally proposed by Markowitz (1952), forms the foundation of modern portfolio theory. The approach aims to construct a portfolio that maximizes expected return for a given level of risk, or equivalently minimizes risk for a given expected return. Risk is quantified as the portfolio’s variance, which accounts for the variability of asset returns and their covariances. The mean-variance framework assumes that investors are rational and risk-averse, preferring portfolios that offer higher returns with lower risk.

Mathematically, the classical optimization problem is expressed as:
\[
\max_{\mathbf{w}} \quad \boldsymbol{\mu}^\top \mathbf{w}
- \frac{\lambda}{2} \mathbf{w}^\top \Sigma \mathbf{w},
\]
subject to
\[
\sum_{i=1}^{n} w_i = 1, 
\qquad 
w_i \ge 0 \quad \forall i.
\]

In this study, the standard mean–variance formulation is extended to incorporate several practical portfolio constraints motivated by real-world execution frictions and institutional guidelines. Each day, the portfolio is rebalanced by solving a constrained optimization problem that reflects limits on turnover, per-asset weights, and transaction costs.

First, the portfolio must remain fully invested and long-only:
\[
\sum_{i=1}^{n} w_i = 1,
\qquad 
w_i \ge 0.
\]

To avoid excessive concentration in any single asset, a common requirement in institutional crypto portfolios, a weight cap is imposed:
\[
w_i \le w_{\max}, \qquad \forall i,
\]
where \( w_{\max} = 0.4 \) in the empirical experiments.

In addition to weight caps, we impose a turnover constraint to prevent excessively frequent trading. Let \( \mathbf{w}_{t-1} \) denote the portfolio weights before rebalancing and \( \mathbf{w}_{t} \) the weights after rebalancing. The turnover is defined as the \( \ell_{1} \)-norm of the adjustment:
\[
\text{Turnover}_t = \| \mathbf{w}_t - \mathbf{w}_{t-1} \|_{1}.
\]
A maximum turnover level is enforced:
\[
\| \mathbf{w}_t - \mathbf{w}_{t-1} \|_{1} \le \tau_{\max},
\]
where \( \tau_{\max} = 0.80 \), limiting daily reallocation to at most 80\% of the portfolio’s notional value.

Transaction costs are incorporated when computing the realized out-of-sample return. Let the one-way transaction cost rate be \( c \) (0.1\% in this study). The total cost of rebalancing at time \(t\) is modeled as:
\[
\text{TCost}_t = 
c \cdot \| \mathbf{w}_t - \mathbf{w}_{t-1} \|_{1},
\]
which reflects proportional costs on all traded notional. This cost is subtracted from the portfolio’s raw return to obtain the net return used in the backtest.

Combining these elements, the daily optimization problem solved in this study becomes:
\[
\max_{\mathbf{w}_t} 
\quad 
\boldsymbol{\mu}_t^\top \mathbf{w}_t
- \lambda \, \mathbf{w}_t^\top \Sigma_t \mathbf{w}_t
\]
subject to
\[
\sum_{i=1}^n w_{i,t} = 1, \qquad
w_{i,t} \ge 0, \qquad
w_{i,t} \le w_{\max},
\]
\[
\| \mathbf{w}_t - \mathbf{w}_{t-1} \|_1 \le \tau_{\max}.
\]
Here, $\boldsymbol{\mu}_{t}$ denotes the adjusted expected return vector constructed at each rolling window (Equation~\ref{eq:mu_adjustment_ridge}). The parameter $\lambda$ controls the degree of risk aversion by weighting the contribution of portfolio variance in the objective function. In this study, $\lambda$ is set to 2. Since expected returns and variances operate on different numerical scales, introducing $\lambda$ prevents the optimization from being dominated by either component. Moreover, $\lambda$ reflects the level of risk aversion implied in the strategy: larger values place greater emphasis on variance reduction, whereas smaller values prioritize return seeking. This formulation aligns with the standard mean–variance framework and leads to more stable and interpretable portfolio weights, especially in settings where covariance estimates may be noisy.

Furthermore, this formulation yields a time-varying pair $(\boldsymbol{\mu}_t, \Sigma_t)$ that reflects both historical distributional properties and forward-looking technical–sentiment signals. When incorporated into the constrained mean–variance optimization problem, these components ensure that the resulting portfolio adheres to realistic investment requirements, including long-only restrictions, weight caps, and turnover discipline, while subsequently accounting for execution frictions through transaction-cost adjustments applied to out-of-sample returns.

\subsection{Evaluation}

The performance of the proposed portfolio optimization strategy is assessed through a comprehensive backtesting framework. The evaluation is conducted on historical cryptocurrency price and news data, using a rolling-window approach to simulate daily rebalancing. Specifically, a 180-day training window is employed to estimate the mean returns and covariance matrix, followed by a 1-day test period for portfolio execution. This process is repeated iteratively across the dataset to produce a time series of out-of-sample daily portfolio returns.

To assess investment performance, three key metrics are computed: \textbf{cumulative return}, \textbf{Sharpe ratio}, and \textbf{maximum drawdown}. The \textbf{cumulative return} measures the total percentage gain or loss over the backtesting period, providing an overall indication of profitability. The \textbf{Sharpe ratio} evaluates risk-adjusted performance by dividing the average daily return by the standard deviation of daily returns, annualized using a factor of $\sqrt{252}$ to account for trading days in a year. A higher Sharpe ratio indicates that the strategy generates greater returns per unit of risk. The \textbf{maximum drawdown} measures the largest peak-to-trough decline in the cumulative returns curve, capturing the worst-case loss an investor might experience during the holding period.

To conduct a comprehensive evaluation, the performance of the proposed strategy is compared against a set of baseline models, which include both market-level reference portfolios and simplified signal-driven trading rules. The first group of baselines consists of a Bitcoin (BTC) buy-and-hold portfolio and an equal-weighted cryptocurrency portfolio. Bitcoin is used as a natural reference asset due to its dominant market capitalization and its central role in shaping price dynamics across the broader cryptocurrency market. The equal-weighted portfolio distributes capital uniformly across all assets, including Bitcoin, offering a diversified comparison that avoids concentration in any single cryptocurrency. For both baselines, daily returns are computed from historical closing prices, and their performance is summarized using cumulative return, annualized Sharpe ratio, and maximum drawdown.

Beyond these market-based references, three additional baselines are designed to isolate the incremental contribution of specific predictive components used in the proposed model. The first is a time-series momentum (TSMOM) strategy, which serves as a classical return-predictive baseline. For each asset, the past \(k\)-day return,
\[
r^{(k)}_{t} = \frac{P_{t}}{P_{t-k}} - 1,
\]
is used as the trading signal: assets with positive momentum receive long positions, while negative signals result in zero allocation. The weights are normalized to enforce full investment under a long-only constraint,
\[
w_{i,t} = \frac{\max(r^{(k)}_{i,t},\, 0)}{\sum_{j} \max(r^{(k)}_{j,t},\, 0)}.
\]

The second baseline is a sentiment-only strategy, constructed to assess the value of news-based information in isolation. The third baseline is a technical-indicator-only variant, in which the expected-return adjustment relies solely on technical features, such as RSI and SMA, while omitting sentiment.

Together, these baseline models allow the evaluation to separately examine the effects of sentiment signals, technical indicators, and momentum dynamics, while also measuring performance relative to widely adopted market-level reference portfolios.

\section{Results}\label{result}

\subsection{Coefficient Stability Analysis}

\begin{figure}[!htbp]
    \centering
    \includegraphics[width=0.85\linewidth]{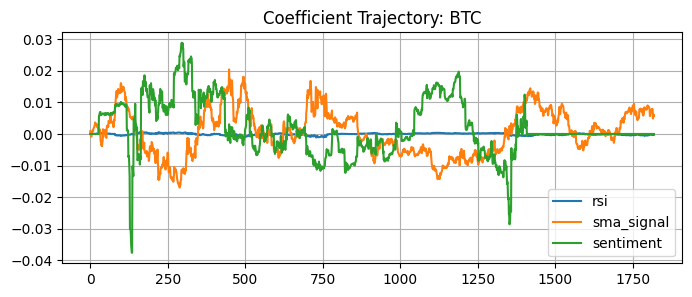}
    \caption{Evolution of estimated Ridge Regression coefficients over time (Bitcoin shown as an illustrative example).}
    \label{fig:btc_coeff_stability}
\end{figure}

Figure~\ref{fig:btc_coeff_stability} illustrates how the coefficients for the Bitcoin evolve over time when they are estimated through a rolling Ridge Regression. The coefficient paths for the other cryptocurrencies follow broadly similar patterns.

The three curves correspond to RSI, the SMA-based signal, and the sentiment measure. As the window moves forward, the coefficient attached to RSI stays very close to zero, drifting only slightly. The SMA coefficient moves in wider ranges, rising and falling in a smoother pattern. The sentiment coefficient is the most active of the three, showing noticeable jumps and reversals, especially during periods of market turbulence. These paths give a sense of how each feature’s influence shifts as market conditions change.

The figure suggests several observations. The near-flat RSI coefficient implies that short-term overbought and oversold patterns offer limited guidance for BTC’s next-day movement in most parts of the sample. Meanwhile, the SMA signal behaves in a more cyclical way, gaining importance during stretches where the market trends cleanly and losing relevance when prices chop sideways. Sentiment behaves quite differently: its coefficient swings more abruptly, which is consistent with how quickly cryptocurrency markets react to news. When the coefficient turns sharply positive, the market is more inclined to incorporate good news into prices; when it becomes negative, it often reflects corrections after overly optimistic sentiment or the impact of adverse news releases. Taken together, the changing coefficients emphasize that the predictive relationships in crypto markets do not remain constant, and that both technical and sentiment-based signals respond to shifting regimes rather than exerting a stable influence over time.

\subsection{Strategy Performance Analysis}

\begin{table}[h]
\centering
\small
\caption{Performance Comparison: Strategies and Baselines}
\label{tab:performance_comparison}
\begin{tabular}{@{}lccc@{}}
\toprule
\textbf{Strategy} &
\makecell{\textbf{Annualized} \\ \textbf{Sharpe Ratio}} &
\makecell{\textbf{Maximum} \\ \textbf{Drawdown (\%)}} &
\makecell{\textbf{Cumulative} \\ \textbf{Return (Final)}} \\
\midrule
Proposed Strategy           & \textbf{0.7102} & \textbf{81.95} & \textbf{6.0555} \\
Proposed (Sentiment Only)   & 0.5976 & 88.99 & 2.6861 \\
Proposed (Technical Only)   & 0.7026 & 83.07 & 4.8510 \\
Momentum                    & 0.6967 & 85.92 & 4.7143 \\
Equal-Weighted Portfolio    & 0.5823 & 75.06 & 2.3911 \\
Bitcoin (BTC) Long-Short    &  0.5454 & 79.63 & 1.9085 \\
\bottomrule
\end{tabular}
\end{table}

\begin{figure}[!htbp]
    \centering
    \includegraphics[width=0.85\linewidth]{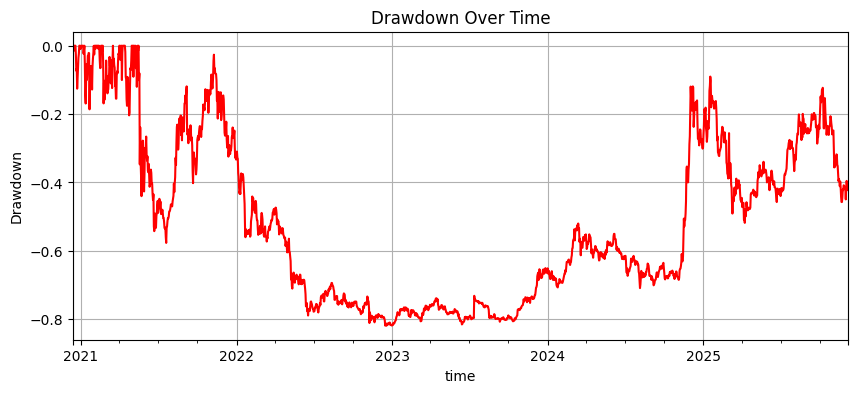}
    \caption{Drawdown of the Proposed Strategy}
    \label{fig: drawdown_plot}
\end{figure}

\begin{figure}[!htbp]
    \centering
    \includegraphics[width=0.85\linewidth]{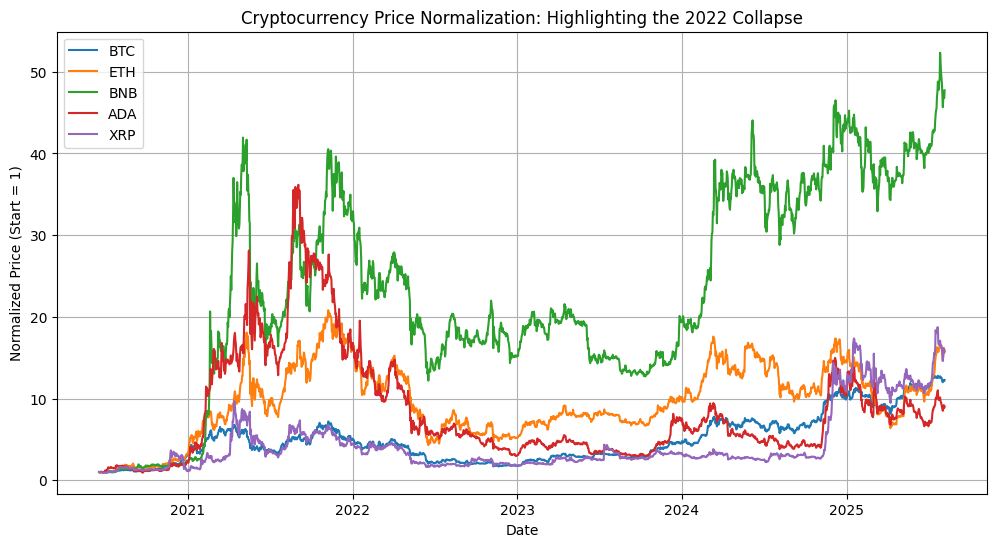}
    \caption{Normalized price trajectories of major cryptocurrencies from 2020 to 2025. 
Each series is scaled by its value at the beginning of the sample, allowing all assets 
to be shown on a comparable scale. A normalized value of 1 represents the initial price, 
and deviations above or below this level reflect proportional gains or losses relative 
to the starting point. This transformation highlights the timing and magnitude of the 
market-wide contraction during the 2022 cryptocurrency downturn.}
    \label{fig: crypto_price_plot}
\end{figure}

\begin{figure}[!htbp]
    \centering
    \includegraphics[width=0.85\linewidth]{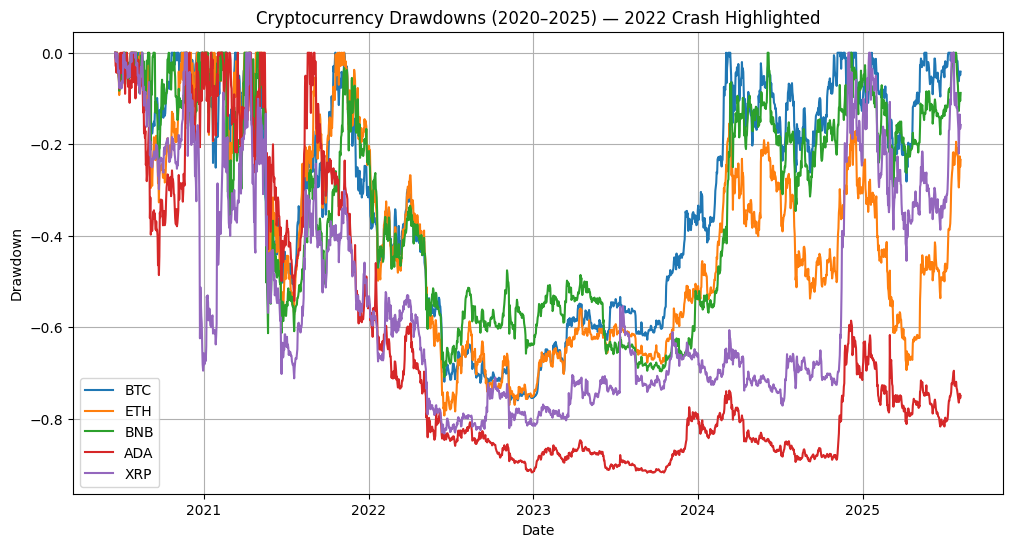}
    \caption{Drawdown of Five Cryptocurrecies from 2020 to 2025}
    \label{fig: crpto_drawdown_plot}
\end{figure}

The results between different strategies are shown in Table~\ref{tab:performance_comparison}. The cumulative return is reported as a raw value rather than a percentage. Comparing the proposed strategy with the benchmark strategies suggests that using both sentiment and technical features together leads to a more balanced outcome than relying on either source alone. Although the sentiment-only strategy benefits from its ability to reflect shifts in market tone and the technical-only strategy captures trend-related movements, each on its own lacks the stability achieved when the signals are combined. The full strategy appears to moderate some of the weaknesses seen in the individual components, producing a more consistent profile across the evaluation metrics.

Furthermore, the momentum strategy produces reasonable returns but still falls short of the proposed strategy in both return level and risk-adjusted performance. The equal-weighted portfolio shows more modest growth compared with the proposed strategy, and although it benefits from diversification, it does not reach the same level of effectiveness. The Bitcoin long-short strategy performs the weakest among all methods in terms of risk-adjusted metrics, and its final return is considerably lower than that of the proposed strategy.

Although the proposed strategy outperforms the benchmark approaches, the results also reveal several weaknesses. The strategy, like the baselines, still experiences sizeable drawdowns, showing that the current signals cannot fully shield the portfolio during periods of rapid market stress. A closer look at the drawdown path helps explain this behavior (Figure~\ref{fig: drawdown_plot}). The portfolio begins to decline sharply around 2022, which corresponds to a broad and prolonged downturn in the cryptocurrency market, marked by major failures within the industry and widespread deleveraging. This pattern is clearly visible in the drawdown chart (Figure~\ref{fig: crpto_drawdown_plot}), where all major assets—BTC, ETH, BNB, ADA, and XRP—enter deep and persistent drawdowns almost simultaneously. A complementary view using normalized price series (Figure~\ref{fig: crypto_price_plot}) shows the same collapse from a different angle: even when all assets are scaled to start at the same value, their trajectories diverge sharply after 2021 and then fall together during 2022, indicating that the drawdown was not isolated to a single coin but was a market-wide phenomenon.

During that period, several high-profile events—including the collapse of Terra–LUNA, liquidity problems at major lending platforms, and the eventual bankruptcy of FTX—triggered severe selling pressure across virtually all digital assets. These shocks caused correlations to spike and erased the diversification normally present in crypto markets, which explains why strategies based on technical, sentiment, or momentum signals struggled to offer meaningful downside protection. The gradual recovery visible in both figures from late 2023 into 2024 aligns with the broader market rebound, suggesting that the model’s behavior is consistent with the evolution of market regimes. While this pattern is expected, it also underscores the need for more refined risk-control methods, improved regime detection, and stronger model robustness to better withstand such extreme market episodes.

\section{Further Discussion: Verifying VADER with Google Gemini}

In addition to using the VADER lexicon-based sentiment analyzer to compute compound scores from news articles, we introduced a verification step based on an LLM to provide a contextual cross-check and actionable investment guidance. Concretely, for each news item we first compute the VADER compound score, and then query the LLM (model version~\texttt{gemini-1.5-flash-latest}) with the article text and the VADER score. The LLM is not fine-tuned because previous research have shown that sometimes LLM without fine-tuning can generate competitive performance compared to LLM fine-tuned with domain knowledge~\cite{Chen2025}.

The prompt used in this study is shown in Table~\ref{tab:llm-prompt}. The prompt is deliberately designed to provide the model with a clear role definition (“You are a crypto investment assistant”) and explicit contextual information, including the news article text and the sentiment score generated by VADER. It then poses three targeted questions, agreement with the sentiment, interpretation, and investment recommendation, ensuring the model delivers reasoning and actionable advice. This structured format helps the model focus on sentiment validation and decision-making, while maintaining interpretability for investors.

\begin{table}[h]
\centering
\caption{Prompt design for Gemini verification of VADER sentiment scores}
\label{tab:llm-prompt}
\begin{tabular}{p{0.95\linewidth}}
\hline
\texttt{You are a crypto investment assistant.}

\texttt{News article:} \\
\texttt{"\{news\_text\}"} \\

\texttt{The VADER sentiment score is: \{vader\_score:.2f\}} \\

\texttt{1. Do you agree with this sentiment?} \\
\texttt{2. What is your interpretation?} \\
\texttt{3. Should the investor increase, decrease, or hold their position?} \\
\texttt{Please give clear reasoning and actionable advice.} \\
\hline
\end{tabular}
\end{table}

The rationale for this two-stage approach is that lexicon methods such as VADER are fast, transparent, and well suited for short-text sentiment aggregation, but they may fail on nuanced language (e.g., sarcasm, contextual domain-specific terms, or disambiguation of named entities). The LLM provides complementary strengths: deeper contextual understanding, the ability to synthesize article-level meaning, and to provide a concise investment recommendation (increase / decrease / hold) together with reasoning. In our pipeline the LLM therefore serves as a secondary verifier and a source of qualitative interpretation rather than as an autonomous trading rule generator.

\begin{table}[h!]
\caption{Sample Gemini response verifying VADER sentiment score}
\label{tab:llm-response-sample}
\begin{tabular}{|p{1.11\linewidth}|}
\hline
\begin{verbatim}
[BTC] News #1: Grayscale Transfers Over 314 BTC Worth $37.5 Million to Coinbase Again
VADER Score: 0.84
Gemini Analysis:

1. **Do I agree with the VADER sentiment score of 0.84?**

Partially. A VADER score of 0.84 indicates strongly positive sentiment. While the news of
Grayscale transferring a large amount of Bitcoin to Coinbase *could* be interpreted positively
(suggesting potential selling or market activity), the accompanying clickbait advertising
("Are You Chasing New Coins? Catch the newest crypto opportunities...") significantly
weakens the positive sentiment. The headline is manipulative and designed to create FOMO
(fear of missing out), which is not a reliable indicator of sound investment strategy.
Therefore, while the core news is potentially positive, the context dilutes the purely positive
sentiment expressed by the VADER score.

2. **Interpretation:**

Grayscale's transfer of Bitcoin to Coinbase is ambiguous. Several interpretations are possible:

* **Selling:** Grayscale might be selling its Bitcoin holdings, potentially impacting the price
  negatively. This is a bearish signal.
* **Trading:** They could be repositioning their assets for trading purposes, buying and selling
  to profit from short-term price fluctuations. This is neutral.
* **Facilitating Transactions:** The Bitcoin could be moved to facilitate other transactions
  or client requests, which is neutral in terms of market impact.

The clickbait advertisement suggests the news source is primarily interested in driving
traffic and promoting altcoins, rather than providing objective financial analysis. This raises
serious questions about the reliability of the source.

3. **Actionable Advice:**

**Hold or cautiously decrease.** Based on the ambiguous nature of the news and the
questionable source, I would *not* advise increasing a Bitcoin position. The possibility of
Grayscale selling their holdings warrants caution. Instead:

* **Ignore the clickbait:** Focus on reputable sources of financial news and analysis, not
  sensationalized headlines designed to elicit impulsive reactions.
* **Conduct thorough research:** Before making any decisions, review broader market trends,
  Grayscale's public statements (if any), and Bitcoin's overall market health. Look for
  confirmation of the transfer's purpose from independent sources.
* **Risk management:** If you are already invested in Bitcoin, consider a small, partial
  decrease in your position as a risk-mitigation strategy if you feel uncomfortable with the
  uncertainty surrounding Grayscale's actions. This would reduce your overall exposure to
  potential negative consequences.
* **Diversification:** Remember that crypto is highly volatile. Diversifying your portfolio
  across different asset classes is crucial to reduce risk. Never put all your eggs in one basket.
\end{verbatim} \\
\hline
\end{tabular}
\end{table}

Table~\ref{tab:llm-response-sample} shows a response sample made by Gemini. In the sample response, Gemini’s reasoning process begins with a direct evaluation of the VADER sentiment score, explicitly stating agreement or disagreement and explaining any nuances behind the stance. It then interprets the underlying news event by outlining multiple plausible scenarios, both positive and negative, and assesses their potential market impacts. Finally, it delivers structured, actionable investment advice that integrates sentiment evaluation, source credibility assessment, market context, and risk management principles. This layered reasoning ensures that recommendations are not based solely on sentiment polarity but also consider the reliability of the information and the broader market environment.

\section{Conclusion and Limitations}\label{sec5}

This study proposed a dynamic cryptocurrency portfolio optimization strategy that integrates technical indicators with sentiment analysis derived from the VADER model. By incorporating both market momentum signals and news-based sentiment, the approach adjusts expected returns prior to portfolio optimization, enabling adaptive allocation decisions in a highly volatile environment. 

The backtest results demonstrate that the proposed strategy consistently outperforms traditional benchmarks, including Bitcoin buy-and-hold, equal-weighted, and momentum-based portfolios, in both cumulative returns and risk-adjusted performance. Integrating sentiment and technical signals produces a more balanced and resilient portfolio, with each component contributing complementary value. Although classical strategies achieve reasonable results, they generally fall short due to limited directional guidance. In general, the findings underscore the potential of combining diverse predictive information to enhance portfolio performance in a robust and practical manner.

The model setup of this study also comes with several limitations. Although the proposed strategy shows improvements over the benchmarks, all strategies, including the proposed one, experience relatively large drawdowns, indicating that the signals used may not fully protect the portfolio during sharp market declines. This suggests limited responsiveness to extreme fluctuations, especially in highly volatile environments. Moreover, the risk model itself relies on a 180-day rolling covariance without shrinkage or robust adjustments, making it sensitive to noise and large price swings. Furthermore, it does not account for heavy-tailed return behavior, which may lead to underestimating extreme events. These limitations highlight opportunities for future improvements in risk-control design, regime awareness, and overall model robustness.

In addition, Google Gemini is used to further validate and interpret the news and news sentiment predicted by VADER. The LLM provides an additional reasoning layer by critically evaluating the sentiment score, considering multiple market scenarios, assessing the credibility of news sources, and delivering structured, risk-aware investment recommendations.

For future research, alternative prompt engineering strategies could be investigated to enhance the quality and consistency of LLM-generated investment recommendations. Comparative studies across different LLM architectures and sizes may reveal performance trade-offs in sentiment interpretation, reasoning depth, and market prediction accuracy. Furthermore, deeper integration between LLM outputs and portfolio optimization frameworks could be explored, such as directly incorporating LLM-derived sentiment scores, scenario analyses, or probability forecasts into the optimization constraints and objective functions. This could enable a more seamless, adaptive decision-making pipeline that leverages the full reasoning capabilities of LLMs in real-time trading environments.

\bibliographystyle{unsrt}  
\bibliography{references}

@article{Hutto_Gilbert_2014, title={VADER: A Parsimonious Rule-Based Model for Sentiment Analysis of Social Media Text}, volume={8}, url={https://ojs.aaai.org/index.php/ICWSM/article/view/14550}, DOI={10.1609/icwsm.v8i1.14550}, number={1}, journal={Proceedings of the International AAAI Conference on Web and Social Media}, author={Hutto, C. and Gilbert, Eric}, year={2014}, month={May}, pages={216-225} }

@Article{math13101693,
  AUTHOR = {{\c{S}}erban, Florentin and Vr{\^\i}nceanu, Bogdan-Petru},
  TITLE = {Mean--Variance--Entropy Framework for Cryptocurrency Portfolio Optimization},
  JOURNAL = {Mathematics},
  VOLUME = {13},
  YEAR = {2025},
  NUMBER = {10},
  ARTICLE-NUMBER = {1693},
  URL = {https://www.mdpi.com/2227-7390/13/10/1693},
  ISSN = {2227-7390},
  DOI = {10.3390/math13101693}
}

@article{BRAUNEIS2019259,
title = {Cryptocurrency-portfolios in a mean-variance framework},
journal = {Finance Research Letters},
volume = {28},
pages = {259-264},
year = {2019},
issn = {1544-6123},
doi = {https://doi.org/10.1016/j.frl.2018.05.008},
url = {https://www.sciencedirect.com/science/article/pii/S1544612318300990},
author = {Alexander Brauneis and Roland Mestel}
}

@Article{jrfm18020077,
AUTHOR = {Gaied Chortane, Sana and Naoui, Kamel},
TITLE = {The End of Mean-Variance? Tsallis Entropy Revolutionises Portfolio Optimisation in Cryptocurrencies},
JOURNAL = {Journal of Risk and Financial Management},
VOLUME = {18},
YEAR = {2025},
NUMBER = {2},
ARTICLE-NUMBER = {77},
URL = {https://www.mdpi.com/1911-8074/18/2/77},
ISSN = {1911-8074},
DOI = {10.3390/jrfm18020077}
}

@ARTICLE{10926863,
  author={Alidaee, Bahram and Wang, Haibo and Wang, Wendy},
  journal={IEEE Access}, 
  title={Comparative Study of Portfolio Optimization Models for Cryptocurrency and Stock Markets}, 
  year={2025},
  volume={13},
  number={},
  pages={49861-49873},
  doi={10.1109/ACCESS.2025.3551628}}

@article{arslan2025bitcoin,
  author    = {Arslan, S.},
  title     = {Bitcoin Price Prediction Using Sentiment Analysis and Empirical Mode Decomposition},
  journal   = {Computational Economics},
  volume    = {65},
  pages     = {2227--2248},
  year      = {2025},
  doi       = {10.1007/s10614-024-10588-3},
  url       = {https://doi.org/10.1007/s10614-024-10588-3}
}

@ARTICLE{10947713,
  author={Shang, Lei},
  journal={IEEE Access}, 
  title={Sentiment-Driven Bitcoin Price Range Forecasting: Enhancing CART Decision Trees With High-Dimensional Indicators and Twitter Dynamics}, 
  year={2025},
  volume={13},
  number={},
  pages={60508-60518},
  doi={10.1109/ACCESS.2025.3557186}}

@INPROCEEDINGS{10825946,
  author={Chen, Qizhao and Kawashima, Hiroaki},
  booktitle={2024 IEEE International Conference on Big Data (BigData)}, 
  title={Stock Price Prediction Using LLM-Based Sentiment Analysis}, 
  year={2024},
  volume={},
  number={},
  pages={4846-4853},
  doi={10.1109/BigData62323.2024.10825946}}

@article{Alghamdi2022,
title = {Cryptocurrency Price Prediction using Forecasting and Sentiment Analysis},
journal = {International Journal of Advanced Computer Science and Applications},
doi = {10.14569/IJACSA.2022.01310105},
url = {http://dx.doi.org/10.14569/IJACSA.2022.01310105},
year = {2022},
publisher = {The Science and Information Organization},
volume = {13},
number = {10},
author = {Shaimaa Alghamdi and Sara Alqethami and Tahani Alsubait and Hosam Alhakami}
}

@Article{bdcc7030137,
AUTHOR = {Frohmann, Markus and Karner, Manuel and Khudoyan, Said and Wagner, Robert and Schedl, Markus},
TITLE = {Predicting the Price of Bitcoin Using Sentiment-Enriched Time Series Forecasting},
JOURNAL = {Big Data and Cognitive Computing},
VOLUME = {7},
YEAR = {2023},
NUMBER = {3},
ARTICLE-NUMBER = {137},
URL = {https://www.mdpi.com/2504-2289/7/3/137},
ISSN = {2504-2289},
DOI = {10.3390/bdcc7030137}
}

@Article{colasanto2022bert,
  author    = {Colasanto, F. and Grilli, L. and Santoro, D. and others},
  title     = {{BERT}'s sentiment score for portfolio optimization: a fine-tuned view in {Black} and {Litterman} model},
  journal   = {Neural Computing and Applications},
  year      = {2022},
  volume    = {34},
  pages     = {17507--17521},
  doi       = {10.1007/s00521-022-07403-1},
  url       = {https://doi.org/10.1007/s00521-022-07403-1}
}

@ARTICLE{10977954,
  author={Taheripour, Esmaeil and Jafar Sadjadi, Seyed and Amiri, Babak},
  journal={IEEE Access}, 
  title={A Novel Approach to Portfolio Construction: An Application of FinBERT Sentiment Analysis and Credibilistic CVaR Criterion}, 
  year={2025},
  volume={13},
  number={},
  pages={76775-76795},
  doi={10.1109/ACCESS.2025.3564615}}

@INPROCEEDINGS{8489208,
  author={Vargas, Manuel R. and dos Anjos, Carlos E. M. and Bichara, Gustavo L. G. and Evsukoff, Alexandre G.},
  booktitle={2018 International Joint Conference on Neural Networks (IJCNN)}, 
  title={Deep Leaming for Stock Market Prediction Using Technical Indicators and Financial News Articles}, 
  year={2018},
  volume={},
  number={},
  pages={1-8},
  doi={10.1109/IJCNN.2018.8489208}}

@INPROCEEDINGS{10284633,
  author={Vincent and Saputra, Benecia Jean and Izzatunnisa, Salsabila Yuka and Lucky, Henry and Iswanto, Irene Anindaputri},
  booktitle={2023 4th International Conference on Artificial Intelligence and Data Sciences (AiDAS)}, 
  title={Stock Market Prediction System Using LSTM with Technical Indicators as Voters}, 
  year={2023},
  volume={},
  number={},
  pages={229-234},
  doi={10.1109/AiDAS60501.2023.10284633}}

@Article{data10040050,
AUTHOR = {Lupu, Radu and Donoiu, Paul Cristian},
TITLE = {Sentiment Matters for Cryptocurrencies: Evidence from Tweets},
JOURNAL = {Data},
VOLUME = {10},
YEAR = {2025},
NUMBER = {4},
ARTICLE-NUMBER = {50},
URL = {https://www.mdpi.com/2306-5729/10/4/50},
ISSN = {2306-5729},
DOI = {10.3390/data10040050}
}

@inproceedings{Chen2025,
  title={Stock Price Change Prediction Using Prompt-Based LLMs with RL-Enhanced Post-Hoc Adjustments},
  author={Qizhao Chen},
  year={2025},
  booktitle={Proceedings of the 2025 4th International Conference on Bigdata Blockchain and Economy Management (ICBBEM 2025)},
  pages={475-483},
  issn={1951-6851},
  isbn={978-94-6463-742-7},
  url={https://doi.org/10.2991/978-94-6463-742-7_46},
  doi={10.2991/978-94-6463-742-7_46},
  publisher={Atlantis Press}
}

@misc{chen2025sentimentawarestockpriceprediction,
      title={Sentiment-Aware Stock Price Prediction with Transformer and LLM-Generated Formulaic Alpha}, 
      author={Qizhao Chen and Hiroaki Kawashima},
      year={2025},
      eprint={2508.04975},
      archivePrefix={arXiv},
      primaryClass={cs.CE},
      url={https://arxiv.org/abs/2508.04975}, 
}

@article{Slusarczyk2025,
author = {{\'S}lusarczyk, D. and {\'S}lepaczuk, R.},
title = {Optimal Markowitz portfolio using returns forecasted with time series and machine learning models},
journal = {J Big Data},
volume = {12},
pages = {127},
year = {2025},
doi = {10.1186/s40537-025-01164-z},
url = {https://doi.org/10.1186/s40537-025-01164-z}
}

@misc{Sakowski2024,
author = {Sakowski, Pawe{\l} and Wysocki, Maciej},
title = {Investment Portfolio Optimization Based on Modern Portfolio Theory and Deep Learning Models},
year = {2024},
month = feb,
howpublished = {\url{https://ssrn.com/abstract=5290128} or \url{http://dx.doi.org/10.2139/ssrn.5290128}},
}

@article{chen2025novel,
  title={A novel sentiment correlation-based method with dual transformer model for stock price prediction},
  author={Chen, Q. and Kawashima, H.},
  journal={International Journal of Data Science and Analytics},
  volume={21},
  number={1},
  year={2025},
  doi={10.1007/s41060-025-00932-7}
}

@article{10.1145/3649451,
author = {Du, Kelvin and Xing, Frank and Mao, Rui and Cambria, Erik},
title = {Financial Sentiment Analysis: Techniques and Applications},
year = {2024},
issue_date = {September 2024},
publisher = {Association for Computing Machinery},
address = {New York, NY, USA},
volume = {56},
number = {9},
issn = {0360-0300},
url = {https://doi.org/10.1145/3649451},
doi = {10.1145/3649451},
journal = {ACM Comput. Surv.},
month = apr,
articleno = {220},
numpages = {42}
}

@INPROCEEDINGS{10973330,
  author={Ke, Zongwu and Zhang, Zeyu and Zhou, Long and Wu, Minqi and Zhang, Tao},
  booktitle={2024 IEEE/WIC International Conference on Web Intelligence and Intelligent Agent Technology (WI-IAT)}, 
  title={Qwen-Fin: Chinese Financial Sentiment Analysis Based on Fine-Tuned Large Language Model}, 
  year={2024},
  volume={},
  number={},
  pages={812-817},
  doi={10.1109/WI-IAT62293.2024.00132}}

@article{10.1093/rfs/hhp063,
    author = {Rapach, David E. and Strauss, Jack K. and Zhou, Guofu},
    title = {Out-of-Sample Equity Premium Prediction: Combination Forecasts and Links to the Real Economy},
    journal = {The Review of Financial Studies},
    volume = {23},
    number = {2},
    pages = {821-862},
    year = {2009},
    month = {08},
    issn = {0893-9454},
    doi = {10.1093/rfs/hhp063},
    url = {https://doi.org/10.1093/rfs/hhp063},
    eprint = {https://academic.oup.com/rfs/article-pdf/23/2/821/24436245/hhp063.pdf},
}

@inproceedings{10.1145/3770177.3770282,
author = {Zhang, Haoyu},
title = {Predicting Future Stock Prices Using Historical Data: A Machine Learning Approach},
year = {2025},
isbn = {9798400720109},
publisher = {Association for Computing Machinery},
address = {New York, NY, USA},
url = {https://doi.org/10.1145/3770177.3770282},
doi = {10.1145/3770177.3770282},
booktitle = {Proceedings of the 2025 International Conference on Economic Management and Big Data Application},
pages = {644–649},
numpages = {6},
location = {
},
series = {ICEMBDA '25}
}

\end{document}